\documentclass[twocolumn,showpacs,superscriptaddress,amsmath,amssymb]{revtex4}
\usepackage{graphicx}% Include figure files
\usepackage{dcolumn}% Align table columns on decimal point
\usepackage{bm}% bold math
\usepackage{color}
\def\bd{
\begin{document}} \def\ed{\end{document}}
\def\bmp{\begin{minipage}} \def\emp{\end{minipage}}
\def\bcc{\begin{center}} \def\ecc{\end{center}}     \def\npg{\newpage}
\def\beq{\begin{equation}} \def\eeq{\end{equation}} \def\hph{\hphantom}
\def\be{\begin{equation}} \def\ee{\end{equation}} \def\r#1{$^{[#1]}$}
\def\n{\noindent} \def\ni{\noindent} \def\pa{\parindent}
\def\hs{\hskip} \def\vs{\vskip} \def\hf{\hfill} \def\ej{\vfill\eject}
\def\cl{\centerline} \def\ob{\obeylines}  \def\ls{\leftskip}
\def\underbar#1{$\setbox0=\hbox{#1} \dp0=1.5pt \mathsurround=0pt
   \underline{\box0}$}   \def\ub{\underbar}    \def\ul{\underline}
\def\f{\left} \def\g{\right} \def\e{{\rm e}} \def\o{\over} \def\d{{\rm d}}
\def\vf{\varphi} \def\pl{\partial} \def\cov{{\rm cov}} \def\ch{{\rm ch}}
\def\la{\langle} \def\ra{\rangle} \def\EE{e$^+$e$^-$} \def\pt{p_{\rm T}}
\def\pti{p_{{\rm T},i}} \def\yti{y_{{\rm T},i}}
\def\ptj{p_{{\rm T},j}}\def\mt{m_{\rm T}} \def\yt{y_{\rm T}} \def\vt{v_{\rm T}}

\def\bitz{\begin{itemize}} \def\eitz{\end{itemize}}
\def\btbl{\begin{tabular}} \def\etbl{\end{tabular}}
\def\btbb{\begin{tabbing}} \def\etbb{\end{tabbing}}
\def\beqar{\begin{eqnarray}} \def\eeqar{\end{eqnarray}}
\def\\{\hfill\break} \def\dit{\item{-}} \def\i{\item}
\def\bbb{\begin{thebibliography}{9} \itemsep=-1mm}
\def\ebb{\end{thebibliography}} \def\bb{\bibitem}
\def\bpic{\begin{picture}(260,240)} \def\epic{\end{picture}}
\def\akgt{\cl{\bf ACKNOWLEDGMENTS}}
\def\fgn{\noindent{\bf\large\bf figure captions}}
%%%%%%%%%%%%%%%%%%%%%%%%%%%%%%%%%%%%%%%%%%%%%%%%%%%%%%%%%%%%%%%%%%%%%%
\def\m1{\langle N_p\rangle} \def\u2{\langle N_{\bar p}\rangle} \def\Nap{N_{\bar
p}}
%%%%%%%%%%%%%%%%%%%%%%%%%%%%%%%%%%%%%%%%%%%%%%%%%%%%%%%%%%%%%%%%%%%%%%%%%%%%%%
\def\lan{\langle}
\def\ran{\rangle}
\def\p{\pi}
\def\ifmath#1{\relax\ifmmode #1\else $#1$\fi}%
\def\rc{\ifmath{{\mathrm{c}}}}
\def\cut{\ifmath{{\mathrm{cut}}}}
\def\rF{\ifmath{{\mathrm{F}}}}
\def\rK{\ifmath{{\mathrm{K}}}}
\def\rp{\ifmath{{\mathrm{p}}}}
\def\rt{\ifmath{{\mathrm{t}}}}
\def\LAB{\ifmath{{\mathrm{LAB}}}}
\def\cut{\ifmath{{\mathrm{cut}}}}
\def\beq{\begin{equation}}
\def\eeq{\end{equation}}

\newcommand{\cinst}[2]{$^{\mathrm{#1}}$~#2\par}
\newcommand{\crefi}[1]{$^{\mathrm{#1}}$}
\newcommand{\crefii}[2]{$^{\mathrm{#1,#2}}$}
\newcommand{\crefiii}[3]{$^{\mathrm{#1,#2,#3}}$}
\newcommand{\HRule}{\rule{0.5\linewidth}{0.5mm}}

\bd
\title{Measurement of anisotropic radial flow in relativistic heavy ion collisions}

\author{Li Lin}
\affiliation{Key Laboratory of Quark and Lepton Physics (MOE)  and 
Institute of Particle Physics, Central China Normal University, Wuhan 430079, China}
\author{Li Na}
\affiliation{Hua-Zhong University of Science and Technology, 430074, China}
\author{Wu Yuanfang} 
\affiliation{Key Laboratory of Quark and Lepton Physics (MOE) and 
Institute of Particle Physics, Central China Normal University, Wuhan 430079, China}

\begin{abstract}
We suggest the azimuthal distribution of mean transverse (radial) rapidity of the
final state particles as a more direct measure of the transverse motion of the source than 
the standard azimuthal multiplicity distribution. 
Using a sample generated by the AMPT model with string melting, we demonstrate that the
azimuthal amplitude of the suggested distribution characterizes the anisotropic radial flow,
and coincides with the parameter of anisotropic radial rapidity extracted from a
generalized blast-wave parametrization.
\end{abstract}


\maketitle
\section{Introduction}

Relativistic heavy ion collisions provide a way to study
the properties of strongly interacting matter. 
The observation of large elliptic flow at RHIC is considered one of the most 
important signatures for the formation of 
the strongly interacting Quark Gluon Plasma
(sQGP)~\cite{qgp,gulassys}. The flow harmonics are Fourier coefficients  
of the azimuthal multiplicity distribution of final state hadrons~\cite{v2}. They are 
considered sensitive probes of the evolution of the
system formed in relativistic heavy ion collisions~\cite{Arthur}.

One common feature of flow harmonics is their mass ordering in the 
low transverse momentum region~\cite{Arthur}. This phenomena can be well
understood by hydrodynamics with a set of kinetic freeze-out
constraints, i.e., the temperature, the radial flow, and the source deformation~\cite{Heinz-PLB}. 
The radial flow is usually described by 2 parameters. 
The first is the isotropic radial velocity (or rapidity, $\vt=\tanh\yt$).  
It presents the surface profile of isotropic transverse expansion of
the source at kinetic freeze-out. 

The other parameter is the anisotropic radial velocity (i.e., the azimuthal dependent radial
velocity). It measures the difference of the radial flow strength in and out of the reaction plane. 
It is introduced to account for the anisotropic radial flow field which arises in {\it
non-central} collisions. The observed elliptic flow can be generated by anisotropic radial flow~\cite{Heinz-PLB,Lisa}. 
Moreover, the shear tension of viscous in hydrodynamics is supposed to be 
proportional to the gradient of radial velocity along the azimuthal direction~\cite{Landao}, 
which is directly related to anisotropic radial velocity. 
The proportionality constant is the shear viscosity. 

In hydrodynamic model~\cite{Kolb, Huovinen, Huichao Song}, 
these parameters are not independent. They are 
related by the initial conditions and the equation of state and their
determination is crucial for theoretical calculations.

The azimuthal distribution of the mean transverse rapidity of final state hadrons 
($\la\yt(\phi)\ra$) directly measures the transverse motion of the source at kinetic freeze-out~\cite{lilin-cpc}.
It should be helpful in determining the parameters of the anisotropic radial rapidity. 
In contrast to the azimuthal multiplicity distribution,
where only the number of particles is concerned, 
here the average is over all particles in a given azimuthal direction.
The influence of the number of particles is excluded. 

As we know, $\la\yt(\phi)\ra$ should contain three parts: average isotropic radial rapidity,
average anisotropic radial rapidity, and average thermal motion rapidity~\cite{Arthur}. 
Since both thermal and radial motions contribute to the isotropic rapidity of the distribution, 
the isotropic radial rapidity itself can not be directly obtained from the distribution.  
So conventionally, the radial flow parameters are extracted from the $\pt$ spectra of the
hadrons~\cite{Schnedermann,Broniowski}, or dileptons~\cite{Jajati,Mohanty}, 
by generalized blast-wave parametrization\cite{Heinz-PLB,Lisa}.

%The azimuthal dependence of the radial flow rapidity 
Fortunately, the thermal motion is isotropic. 
As such, it does not contribute to the anisotropic radial flow. The azimuthal amplitude 
of the mean transverse rapidity distribution should correspond directly to the anisotropic
radial rapidity. It is interesting to see the features of the azimuthal distribution
of mean transverse rapidity, and how its azimuthal amplitude relates
to the parameters of anisotropic radial rapidity extracted 
by a generalized blast-wave parametrization.

In the paper, we define the $\la\yt(\phi)\ra$ in section II. 
Using a sample generated by the AMPT model with string melting
~\cite{ampt1,ampt2}, the suggested distribution 
and its particle mass and centrality dependence are presented. 
These show that the isotropic and anisotropic parts of the suggested distribution behave  
as the expected radial flow (with a random thermal component), and anisotropic radial flow, respectively. 
In section III, the $\pt$ spectra of 6 particle species and their corresponding elliptic flows, $v_2(\pt)$, 
are presented. Fitting these spectra and elliptic flows 
by a generalized blast-wave parametrization, 
the temperature, and the radial flow parameters are obtained. It
is found that the parameter of anisotropic radial rapidity is well
described by the azimuthal amplitude of the suggested distribution. 
Finally, the summary and conclusions are given in section IV.

\section{Azimuthal distribution of mean transverse rapidity}

Usually, the transverse rapidity of a final state hadron is considered a
good approximation of its transverse rapidity at kinetic freeze-out~\cite{nature}.
It is defined as,

\begin{equation}
\yt=\ln(\frac{\mt+\pt}{m_{0}})
\end{equation}

\noindent where $m_{0}$ is the particle mass in the
rest frame, $\pt$ is transverse momentum, and
$\mt=\sqrt{m_0^2+\pt^2}$ is the transverse mass.
The mean transverse rapidity in a given azimuthal angle bin
is defined as the summation of all particles' rapidities 
divided by the total number of particles, i.e., 
\begin{equation}
\la\yt(\phi-\psi_r)\ra=\frac{1}{N_{\mathrm{event}}}\sum_{e=1}^{N_{\mathrm{event}}}\frac{1}{N^e_m}\sum_{i=1}^{N^e_{m}}\yti^e(\phi_m-\psi_r),
\end{equation}

%\begin{equation}
%\la\yt^e(\phi_m-\psi_r)\ra=\frac{1}{N^e_m}\sum_{i=1}^{N_{m}}\yti^e(\phi_m-\psi_r),
%\end{equation}

\noindent where $y_{T,i}^e$ is the transverse rapidity of the $i$th
particle and $N_m^e$ is the 
total number of particles in $m$th azimuthal angle bin in $e$th event. 
The direction of reaction plane is given by $\psi_r$, which is zero in model calculation, and can be determined
in experiment by 3 similar ways as that for the azimuthal multiplicity distribution~\cite{arthur}.  
By definition, the influence from the number of particles is therefore removed.
Eq.~(2) measures the mean transverse motion in azimuthal direction~\cite{lilin-cpc}.

%In experimental data analysis, the azimuthal direction is respect to the direction
% of reaction plane, $\psi_r$. The reaction plane is formed by impact parameter and beam direction.
% It has been carefully determined event by event in azimuthal multiplicity
% distribution~\cite{Arthur}. We can approximately use it in the
%azimuthal distribution of transverse rapidity at first step.
% It can also be determined similarly as that in azimuthal multiplicity distribution.
% In this paper, we focus our study on the model.
%It is zero in the model analysis.

In order to see the generic features of the defined distribution, 
we present the distribution from a sample of Au + Au
collisions at 200 GeV generated  
by the AMPT model with string melting~\cite{ampt2} in Fig.~1(a). 
For the following comparison, we use a longitudinal rapidity window of $|y_{\rm L}|<0.1$, 
the same as published spectra data from the STAR experiment~\cite{STAR-v2,Abelev}.

\begin{figure*}
\includegraphics[width=7.2in]{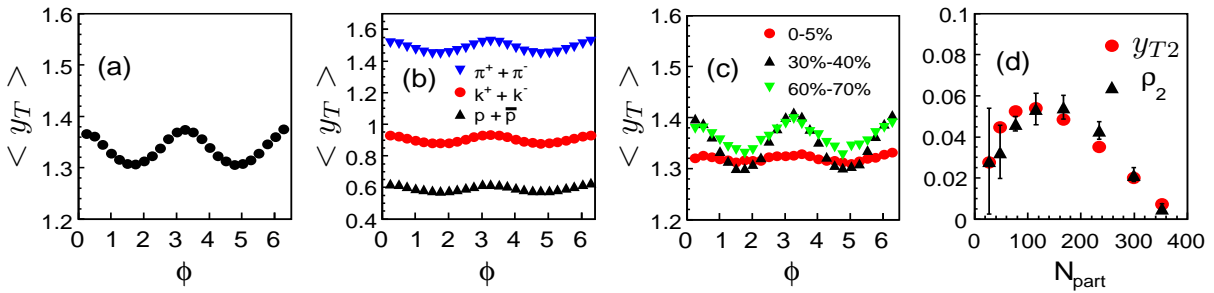}
\caption{\label{Fig. 1}(Color online) (a) $\la\yt(\phi)\ra$ of minimum bias sample,
(b)$\la\yt(\phi)\ra$ for 3 different mass particles,
(c) $\la\yt(\phi)\ra$ at three
different centralities, $0-5\%$ (red points), $30\%-40\%$ (black triangles),
and $60\%-70\%$ (green down triangles), (d) The centrality dependence
of $y_{T2}$ (red points), and $\rho_2$ (black triangles). }
\end{figure*}

From the figure, we can see it is a periodic function of azimuthal angle and can
be well fitted by,
\begin{equation}
\la\yt(\phi)\ra=y_{\rm T0}+y_{\rm T2}\cos(2\phi).
\end{equation}
Eq.~(3) consists of two parts: an isotropic mean rapidity, $y_{\rm T0}=1.3371\pm
0.0001$, and a mean azimuthal dependent rapidity amplitude, $y_{\rm
T2}=0.0334\pm0.0002$. This directly indicates the anisotropic distribution of transverse 
motion, in addition to the known anisotropy distribution of particle number. 
Further, the anisotropic amplitude, $y_{\rm T2}$, should correspond to the
parameter of anisotropic radial rapidity. 

The isotropic part is a combination of radial rapidity and thermal motion rapidity.
As we know, the thermal motion is mainly determined by the temperature and particle mass.
For a system at fixed temperature, the lighter particles should have
larger thermal velocity. To see this feature in isotropic
rapidity we plot the distributions of three different
particle species, $\pi$, $k$, and $p$ in Fig.~1(b). It shows that the lightest
particles (pions) have the largest isotropic rapidities, while the heaviest
particles (protons) have the smallest isotropic rapidities, and 
intermediate mass particles (kaons) have rapidities between them. 
These indicate that their isotropic rapidities are 
ordered as expected from random thermal motion.

To see the influence of centrality on the anisotropic part,
the azimuthal distributions of mean transverse rapidity
at three typical centralities, $0-5\%$, $30\%-40\%$,
and $60\%-70\%$, are presented in Fig~1(c).
From the figure, we can see that the distributions are
azimuthal angle dependent in non-central collisions, i.e.,
the mid-central and peripheral collisions with centralities
of $30\%-40\%$ and $60\%-70\%$. It becomes almost flat and azimuthal
angle independent in central collisions ($0-5\%$).
So the azimuthal dependent part of the distribution appears only in
non-central collisions.

This is consistent with the fact that only two parameters, the
temperature and radial rapidity, are required to describe the observed
$\pt$ spectra in {\rm central} collisions, as done in early
blast-wave parametrization~\cite{Lisa}. However, the parameter of
anisotropic radial flow is necessary for non-central
collisions~\cite{Heinz-PLB,Lisa}.

The centrality dependence of the azimuthal amplitude, $y_{\rm T2}$,  is
shown in Fig.~1(d) by red solid circles. It has a maximum in mid-central collisions,
decreases toward peripheral and central collisions, and is close to zero in central collisions.

The disappearance of $y_{\rm T2}$ in central collisions also indicates that
the thermal motion, which exists in central collisions as well, is isotropic
and does not contribute to the anisotropic radial rapidity.
Therefore, the azimuthal amplitude of the suggested distribution
describes the parameter of anisotropic radial rapidity.
In order to show this quantitatively, we will compare it with 
the parameter that is extracted from the {\it same} sample by a generalized 
blast-wave parametrization in the following section.

\section{The parameters of radial flow }

The blast-wave model is currently the only model that simply includes the radial flow parameters. 
It is motivated from hydrodynamics with the kinetic freeze-out
parameters~\cite{Lisa,Schnedermann,Siemens,STAR-v2,Adams2,blast-v2}.
It is assumed that the longitudinal expansion is boost
invariant~\cite{Bjorken}. The single-particle spectrum is
given by the Cooper-Frye formalism (as in hydrodynamics)~\cite{Cooper},
\begin{equation}
 E\frac{d^3N}{d^3p}\propto\frac{1}{(2\pi)^3}\int_{\Sigma_{f}}{p^{\mu}d\sigma_{\mu}(x)f(x,p)},
\end{equation}
\noindent  where $f(x,p)$ is the
momentum distribution at space-time point $x$.
Eq.~(4) is an integral over a freeze-out hyper-surface, and sums over the
contributions from all space-time points.

Originally, local thermal equilibrium  is assumed to be reached
at kinetic freeze-out and a Boltzman distribution of the momentum
is applied~\cite{Schnedermann}. It has been shown recently that a 
Tsallis distribution provides an even better description for all $\pt$
spectra from elementary to nuclear collisions~\cite{Tsallis,Shao}.
So, we use the Tsallis distribution for $f(x,p)$, i.e.,

\begin{equation}
 f(x,p)=\Biggl[ 1+\frac{q-1}{T(x)}\biggl( p\cdot\\u(x)-\mu(x)\biggr)\Biggr]^{-\frac{1}{q-1}},
\end{equation}

\noindent where $q$ is the parameter characterizing the degree of
non-equilibrium, and $T$ is the kinetic freeze-out temperature.
Thus the transverse momentum spectrum can be given
by~\cite{Tang},

\begin{eqnarray}
\frac{dN}{\pt d\pt d\phi} &\propto &
\int_{0}^{2\pi}d\phi_{s}\int_{-y_{b}}^{y_{b}}dye^{\sqrt{y_{b}^2-y^2}}\cosh
y\int_{0}^{R}\mt rdr
\nonumber\\
& & \ \ \Biggl[ 1+\frac{q-1}{T}\biggl( \mt\cosh y\cosh\rho  \nonumber\\
& & \qquad -\pt\sinh\rho\cos(\phi_{b}-\phi)\biggr)\Biggr] ^{-\frac{1}{q-1}}
\end{eqnarray}

\noindent where $\mt$ and $\pt$ are transverse mass and
transverse momentum of the particle, respectively, and 
$y_{b}=\ln(\sqrt{s_{NN}}/m_{N})$ is the beam rapidity~\cite{Wong}.

According to the generalized blast-wave parametrization, the radial
flow rapidity which controls the magnitude of the transverse expansion
velocity is~\cite{Lisa,Poskanzer,STAR-v2,Yongseok},

\begin{equation}
\rho=\widetilde{r}\bigl(\rho_{0}+\rho_{2}\cos(2\phi_{b})\bigr)
\end{equation}

\noindent where
$\widetilde{r}=\sqrt{(r\cos(\phi_{s})/R_{X})^2+(r\sin(\phi_{s})/R_{Y})^2}$.
$\rho_0$ is the isotropic radial flow rapidity, and $\rho_2$ is the
amplitude of the anisotropic radial flow rapidity, respectively. The
greater the magnitude of $\rho_{2}$, the larger the momentum-space
anisotropy. Here, $\phi_{s}$ is the azimuthal angle in coordinate
space and $\phi_{b}$ is the azimuthal angle of the boost source
element defined with respect to the reaction plane. They are
related by $\tan(\phi_{b})=(R_{X}/R_{Y})^2\tan(\phi_{s})$.

There are 5 undetermined parameters: the temperature ($T$), 
isotropic radial flow rapidity ($\rho_0$) and anisotropic
radial flow rapidity ($\rho_2$), $q$ of the Tsallis distribution, and $R_{X}/R_{Y}$. 
Since all the particles are
assumed to move with a common radial flow velocity, the mean kinetic
freeze-out parameters are usually obtained by the simultaneous fitting of spectra from several hadrons
~\cite{Abelev,Adams2} and elliptic flow~\cite{Lisa}.
Elliptic flow, $v_2(\pt$), is the second coefficient of the Fourier expansion of
azimuthal multiplicity distribution~\cite{Ollitrault1,Voloshin}, and defined as,

\begin{equation}
v_{2}(\pt)=\frac{\int_{-y_{b}}^{y_{b}}dy\int_{0}^{2\pi}d\phi\cos(2\phi)\frac{dN}{\pt
d\pt dy d\phi}}{\int_{-y_{b}}^{y_{b}}dy\int_{0}^{2\pi}d\phi\frac{dN}{\pt
d\pt dy d\phi}}.
\end{equation}

In Fig.~2, the $\pt$ spectra of six particles, $\pi^{\pm}$,
$K^{\pm}$, $\overline{p}$, and $p$, of the {\it same} sample,
are presented by red solid circles. The differential elliptic
flow $v_{2}(\pt)$ of pions, kaons, and protons are presented in
Fig.~3 by black triangles, red solid circles and blue triangles,
respectively. The error bars only include statistical
errors. They are very small in comparison with the experimental
data~\cite{STAR-v2}. Typically, the systematic errors are considered
to be $5\%$ when fitting the simulated data~\cite{Shan}. Due to resonance
decays in the low momentum region of pions~\cite{Abelev}, the data points
in the low $\pt$ regions of the spectra are excluded in this fitting.

\begin{figure}
\includegraphics[width=4.2in]{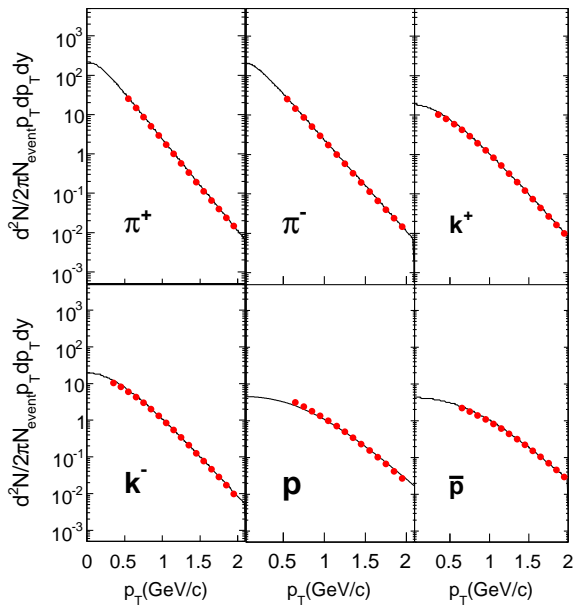}
 \caption{\label{Fig. 3}(Color online)  The transverse momentum spectra for
$\pi^{\pm}$, $K^{\pm}$, $\overline{p}$ and $p$ within $|y_{\rm L}|<0.1$ for
the sample of Au+Au collisions at $\sqrt{s_{NN}}=200$ GeV generated
by the AMPT model with string melting.}
\end{figure}

\begin{figure}
\includegraphics[width=3.8in]{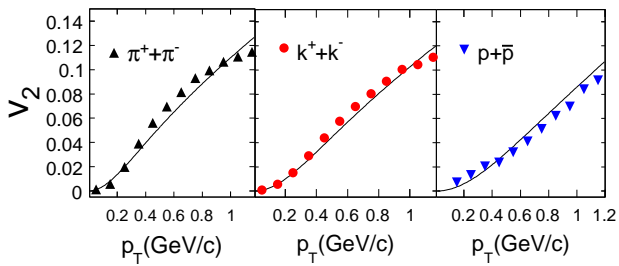}
 \caption{\label{Fig. 3}(Color online) The differential elliptic flow  $v_{2}(p_{T})$
 for different particle species within $|y_{\rm L}|<0.1$ for the sample of 
 Au+Au collisions at $\sqrt{s_{NN}}=200$ GeV
 generated by the AMPT model with string melting.}
\end{figure}

Using Eqs.~(6), and (8), the fitting curves in each plots of Fig.~2
and 3 are drawn. They describe well the corresponding data points of
the $\pt$ spectra and elliptic flow. The
fitting parameters are $T=96.1\pm1.0(MeV)$, $\rho_{0}=0.73\pm0.01$,
and $\rho_{2}=0.035\pm0.003$. This temperature is the same magnitude
as that given by hydrodynamics~\cite{Heinz-PLB,Huichao Song}, and
experimental data~\cite{STAR-v2}. The parameter of anisotropic radial flow
rapidity, $\rho_2$, is very close to the azimuthal amplitude of the suggested 
distribution, $y_{\rm T2}=0.0334\pm0.0002$.

The centrality dependence of $\rho_{2}$ is shown in Fig.~1(d) by black triangles.
We can see that at each centrality, $\rho_2$ is very close to $y_{\rm T2}$.
The azimuthal amplitude of the suggested distribution 
coincides with the parameter of anisotropic radial flow rapidity extracted from a 
generalized blast-wave parametrization.

\begin{figure}
\includegraphics[width=3.5in]{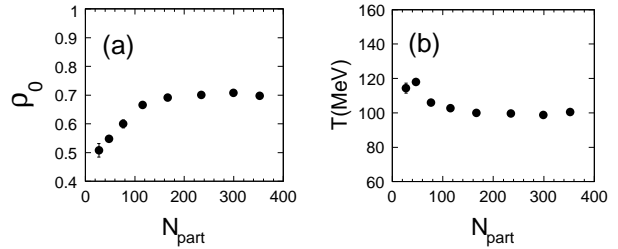}
\caption{\label{Fig. 4} The centrality dependence of (a) the isotropic radial flow rapidity
and (b) the temperature obtained by blast-wave parametrization from the sample of Au+Au collisions at
200 GeV generated by the AMPT model with string melting.}
\end{figure}

To complete the discussion of the fitting, the centrality dependence of $\rho_{0}$  
and $T$  are shown in Figs. 4(a) and (b), respectively. We can see that $\rho_{0}$ increases 
with the increasing of number of participants, and reaches a maximum in central collisions. 
The temperature changes with centrality
in the opposite way as that of radial flow, $\rho_{0}$.
The radial flow and the temperature are negatively correlated, as expected~\cite{Schnedermann,Bearden}.
This is because the produced hadrons in central collisions
have more time to cool down, rescatter, and form a stronger
radial expansion flow. In peripheral collisions, there is not
enough time to convert thermal energy to the collective flow motion.

\section{Summary and conclusions}

In this paper the measurement of the azimuthal distribution
of mean transverse rapidity of final state hadrons is proposed as 
a more direct probe of the transverse motion of the source than 
the known azimuthal multiplicity distribution. 

Using the sample generated by the AMPT model with string melting, we show 
that the isotropic part of the distribution is a combination of the radial and
thermal motions. The azimuthally dependent part measures the anisotropy 
of transverse motion arising from non-central collisions.

Using a generalized blast-wave parametrization, we further extract the temperature and 
radial flow parameters from the same sample. It is found that the parameter of 
anisotropic radial rapidity coincides with the azimuthal amplitude of the 
suggested distribution.

The $\la\yt(\phi)\ra$ provides a
direct measurement of the anisotropic radial rapidity. This is 
important for hydrodynamic calculations and for a direct 
measurement of shear viscosity in relativistic heavy ion collisions~\cite{Meijuan}.

\section{Acknowledgement}

We are grateful for the valuable comments of
Dr. Zhangbu Xu, Fuqiang Wang and Terence Tarnowsky. The first author would thank Dr. Zebo Tang for 
effective helps in using Tsallis distribution. This work is supported in part by the NSFC of 
China with project No. 10835005 and 11221504, and MOE of China with project No. IRT0624 and 
No. B08033.

\ed